\documentclass[runningheads]{llncs}
\usepackage{graphicx}
\usepackage{amsmath,amssymb}
\usepackage{booktabs}
\usepackage{color}
\usepackage{listings}
\usepackage{xcolor}
\usepackage{soul}
\usepackage{url}
\usepackage{multirow}
\usepackage{fontawesome}
\usepackage{nicefrac}
\usepackage{subcaption}
\usepackage[inline]{enumitem}
\usepackage{wrapfig}
\usepackage{wasysym}
\usepackage{footnote}
\usepackage{tikz}
\usetikzlibrary{matrix}
\definecolor{Gray}{gray}{0.9}

\newcommand{\samplefix}{SampleFix}
\newcommand{\dssmaplefix}{DS-SampleFix}

\newcommand{\figref}{Figure}

\newcommand{\equref}{Equation}
\newcommand{\tableref}{Table}

\definecolor{ao}{rgb}{0.0, 0.5, 0.0}
\newcommand{\myparagraph}[1]{\vspace{0.0em}\noindent\textbf{#1.}}

\begin{document}
\title{SampleFix: Learning to Generate Functionally Diverse Fixes}
%
%
\author{Hossein Hajipour\inst{1}\and
Apratim Bhattacharyya\inst{2}\and Cristian-Alexandru Staicu\inst{1}\and 
Mario Fritz\inst{1}}
\authorrunning{H. Hajipour et al.}
%
\institute{CISPA Helmholtz Center for Information Security, Germany \and
Max Planck Institute for Informatics, Germany
}
\maketitle              
\begin{abstract}
Automatic program repair holds the potential of dramatically improving the productivity of programmers during the software development process and correctness of software in general. Recent advances in machine learning, deep learning, and NLP have rekindled the hope to eventually fully automate the process of repairing programs. 
  However, previous approaches that aim to predict a single fix are prone to fail due to uncertainty about the true intend of the programmer. Therefore, we propose a generative model that learns a \textit{distribution} over potential fixes. Our model is formulated as a deep conditional variational autoencoder that can efficiently sample fixes for a given erroneous program. In order to ensure \textit{diverse} solutions, we propose a novel regularizer that encourages diversity over a semantic embedding space. Our evaluations on common programming errors show for the first time the generation of diverse fixes and strong improvements over the state-of-the-art approaches by fixing up to $45\%$ of the erroneous programs. We additionally show that for the $65\%$ of the repaired programs, our approach was able to generate multiple programs with diverse functionalities.  
\end{abstract}
\section{Introduction}

\begin{wrapfigure}[14]{R}{0.4\linewidth}
\vspace{-1.35cm}
\centering
    \includegraphics[width=0.99\linewidth]{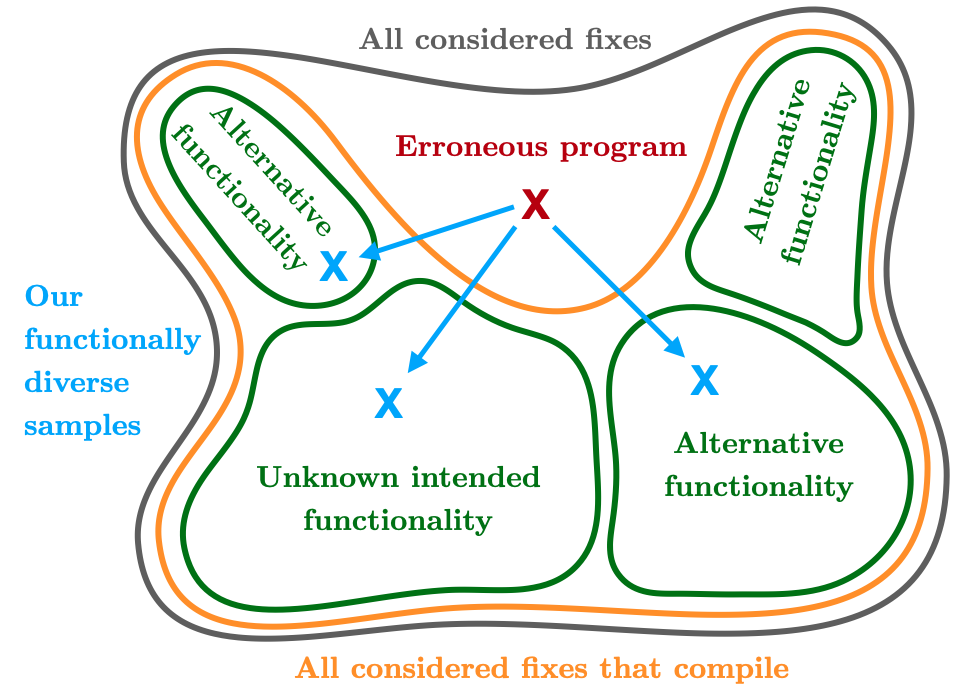}
    \caption{Our SampleFix approach with diversity regularizer promotes sampling of diverse fixes, that account for the inherent uncertainty in the automated debugging task.}
\label{fig:bubbles}
\end{wrapfigure}

Software development is a time-consuming and expensive process. Unfortunately, programs written by humans typically come with bugs, so significant effort needs to be invested to obtain code that is only likely to be correct. Debugging is also typically performed by humans and 
can contain mistakes. This is neither desirable nor acceptable in many critical applications. Therefore, automatically locating and correcting program errors~\cite{GouesPR19} offers the potential to increase productivity as well as improve the correctness of software. 

Advances in deep learning \cite{alexnet,ng11unsupervised}, computer vision \cite{Girshick15fast,Simonyan15}, and NLP \cite{sutskever14seq2seq,Bahdanau2015NeuralMT} have dramatically boosted the machine's ability to automatically learn representations of natural data such as images and natural language contents for various tasks.  Deep learning models also have been successful in learning the distribution over continuous \cite{cvae15sohn,autoencodr14kingma} and discrete data \cite{Maddison2017TheCD,jang18gumbel}, to generate new and diverse data points \cite{three_pillars}. These advances in machine learning and the advent of large corpora of source code~\cite{bigcode} provide new opportunities toward harnessing deep learning methods to understand, generate, or debug programs.

\begin{figure*}[h] 
	\centering
	    \centering
		\includegraphics[width = 0.85\textwidth]{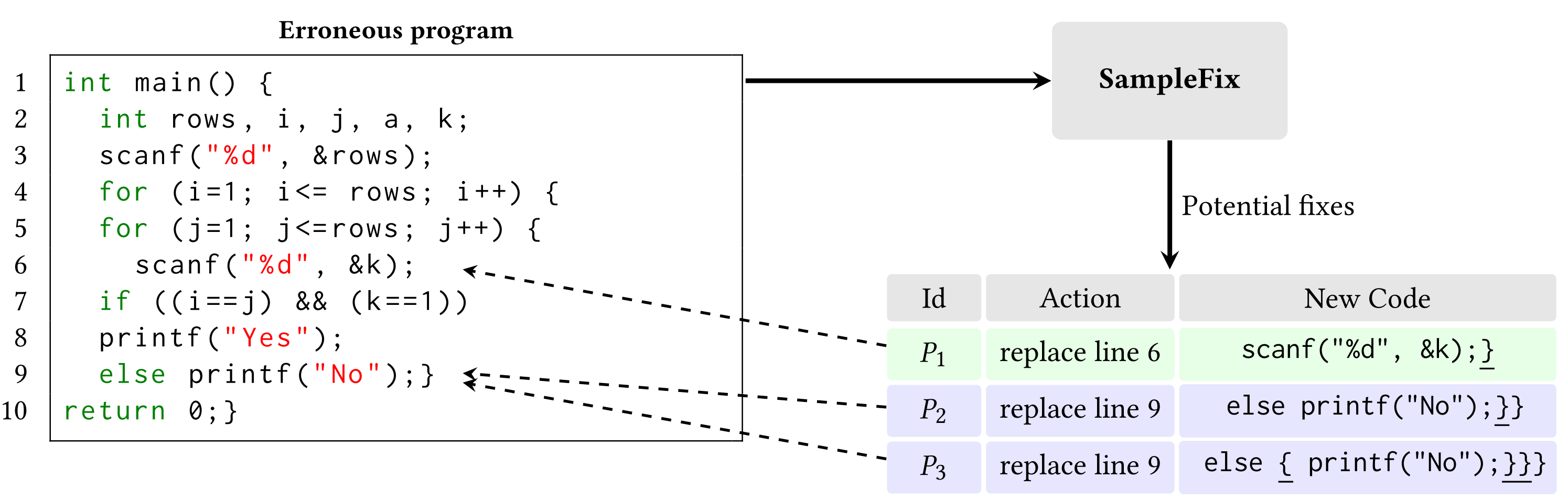}
	
	\caption{SampleFix captures the inherent ambiguity of the possible fixes by sampling multiple potential fixes for the given erroneous real-world program. Potential fixes with the same functionality are highlighted with the same color and the newly added tokens are underlined.}
	\label{fig:teaser}
	\vspace{-0.5cm}
\end{figure*}

Prior works in automatic program repair predominantly rely on expert-designed rules and error models that describe the space of the potential fixes  \cite{singh2013automated,d2016qlose}. Such hand-designed rules and error models are not easily adaptable to the new domains and require a time-consuming process.

In contrast, learning-based approaches provide an opportunity to adapt such models to the new domain of errors. Therefore, there has been an increasing interest to carry over the success stories of deep learning in NLP and related techniques to employ learning-based approaches to tackle the ``common programming errors'' problem \cite{Gupta2017DeepFixFC,gupta2019RLAssist}. Such investigations have included compile-time errors such as missing scope delimiters, adding extraneous symbols, using incompatible operators. Novice programmers and even experienced developers often struggled with these types of errors \cite{seo2014programmers}, which is usually due to lack of attention to the details of programs and/or programmer's inexperience. 

 Recently, Gupta et al. \cite{Gupta2017DeepFixFC} proposed a deep sequence to sequence model called DeepFix where, given an erroneous program, the model predicts the locations of the errors and a potential fix for each predicted location. The problem is formulated as a deterministic task, where the model is trained to predict a single fix for each error. However, different programs -- and therefore also their fixes -- can express the same functionality. Besides, there is also uncertainty about the intention of the programmer.
 \figref \, \ref{fig:bubbles} illustrates the issue. Given an erroneous program (buggy program), there is a large number of programs within a certain edit distance. A subset of these, will result in successful compilation. The remaining programs will still implement different functionalities and -- without additional information or assumptions -- it is impossible to tell which program/functionality was intended.
 In addition, previous work \cite{SmithBGB15} also identified overfitting as one of the major challenges for learning-based automatic program repair. We believe that one of the culprits for this is the poor objectives used in the training process, e.g., training a model to generate a particular target fix.
 
 Let us consider the example in \figref \,\ref{fig:teaser} from the dataset of DeepFix \cite{Gupta2017DeepFixFC}. This example program is incorrect due to the imbalanced number of curly brackets. In a traditional scenario, a compiler would warn the developer about this error. For example, when trying to compile this code with GCC, the compiler terminates with the error ``expected declaration or statement at end of input'', indicating line 10 as the error location. Experienced developers would be able to understand this cryptic message and proceed to fixing the program. Based on their intention, they can decide to add a curly bracket either at line 6 (patch $P_1$) or at line 9 (patch $P_2$). Both these solutions would fix the compilation error in the erroneous program, but the resulting solutions have different semantics. 
 
Hence, we propose a deep generative framework to automatically correct programming errors by learning the distribution of potential fixes. We investigate different solutions to model the distribution of the fixes and sample multiple fixes, including different variants of Conditional Variation Autoencoders (CVAE) and beam search decoding. It turns out (as we will also show in our experiments) CVAE and beam search decoding are complementary, while CVAE is computationally more efficient in comparison to beam search decoding.
Furthermore, we encourage diversity in the candidate fixes through a novel regularizer which penalizes similar fixes for an identical erroneous program and significantly increases the effectiveness of our approach.  The candidate fixes in \figref \,\ref{fig:teaser} are generate by our approach, illustrating its potential for generating both diverse and correct fixes. For a given erroneous program, our approach is capable of generating diverse fixes to resolve the syntax errors.

To summarize, the contributions of this paper are as follows,
\begin{enumerate*}
    \item We propose an efficient generative method to automatically correct common programming errors by learning the distribution over potential fixes.
    \item We propose a novel regularizer to encourage the model to generate diverse fixes.
    \item Our generative model together with the diversity regularizer shows an increase in the diversity and accuracy of fixes, and a strong improvement over the state-of-the-art approaches.
\end{enumerate*}

\section{Related Work}
\label{related_Work}

Our work builds on the general idea of sequence-to-sequence models as well as ideas from neural machine translation. We phrase our approach as a variational auto-encoder and compare it to prior learning-based program repair approaches. We review the related work in order below:

\subsection{Neural Machine Translation}
Sutskever et al. \cite{sutskever14seq2seq} introduces neural machine translation and casts it as a sequence-to-sequence learning problem. The popular encoder-decoder architecture is introduced to map the source sentences into target sentences. One of the major drawbacks of this model is that the sequence encoder needs to compress all of the extracted information into a fixed-length vector. Bahdanau et al. \cite{Bahdanau2015NeuralMT} addresses this issue by using attention mechanism in the encoder-decoder architecture, where it focuses on the most relevant part of encoded information by learning to search over the encoded vector. In our work, we employ a sequence-to-sequence model with attention to parameterize our generative model. This model gets an incorrect program as input and maps it to many potential fixes by drawing samples on the estimated distribution of the fixes.

\subsection{Variational Autoencoders}
\label{rw_variational_autoencoder}
The variational autoencoders \cite{autoencodr14kingma,variational15rezende} is a generative model designed to learn deep directed latent variable based graphical models of large datasets. The model is trained on the data distribution by maximizing the variational lower bound of the log-likelihood as the objective function. Bowman et al.  \cite{gensen16bowman} extend this framework by introducing an RNN-based variational autoencoder to enable the learning of latent variable based generative models on text data. The proposed model is successful in generating diverse and coherent sentences. To model conditional distributions for the structured output representation Sohn et al. \cite{cvae15sohn} extended variational autoencoders by introducing an objective that maximizes the conditional data log-likelihood. In our approach, we employ an RNN-based conditional variational autoencoder to model the distribution of the potential fixes given erroneous programs. Variational autoencoder approaches enable the efficient sampling of accurate and diverse fixes. 

\subsection{Learning-based Program Repair}
\label{rw_program_repair}
Recently there has been a growing interest in using learning-based approaches to automatically repair the programs \cite{monperrus2018automatic}.  Long and
Rinard \cite{long2016automatic} proposed a probabilistic model by designing code features to rank potential fixes for a given program.  Pu et al. \cite{pu2016sk_p} employ an encoder-decoder neural architecture to automatically correct programs. In these works and many learning-based programming repair approaches, enumerative search over programs is required to resolve all errors. However, our proposed framework is capable of predicting the location and potential fixes by passing the whole program to the model. Besides this, unlike our approach, which only generates fixes for the given erroneous program, Pu et al. \cite{pu2016sk_p} need to predict whole program statements to resolve the errors.

There are two important program repair tasks explored in the literature: fixing syntactic errors and fixing semantic ones. While in the current work we propose a technique for fixing syntactic errors, we believe that our observation about the diversity of the fix has implications for the approaches aimed at repairing semantic bugs as well. Most of the recent work in this domain aim to predict a unique fix, often extracted from a real-world repository. For example, Getafix~\cite{BaderSP019}, a recent approach for automatically repairing six types of semantic bugs, is evaluated on a set of 1,268 unique fixes written by developers. Similarly, DLfix~\cite{li2020dlfix} considers a bug to be fixed only if it exactly matches a patch provided by the developer. 
While this is an improved methodology in the spirit of our proposal it is highly dependent on the performance of the test suite oracle which may not always capture the developer's intent.

DeepFix \cite{Gupta2017DeepFixFC}, RLAssist \cite{gupta2019RLAssist}, and DrRepair \cite{yasunaga2020repair} uses neural representations to repair syntax errors in programs. In detail, DeepFix \cite{Gupta2017DeepFixFC} uses a sequence-to-sequence model to directly predict a fix for incorrect programs. In contrast, our generative framework is able to generate multiple fixes by learning the distribution of potential correctness. Therefore, our model does not penalize, but rather encourages diverse fixes. RLAssist \cite{gupta2019RLAssist} repairs the programs by employing a reinforcement learning approach. They train an agent that navigate over the program to locate and resolve the syntax errors. In this work, they only address the typographic errors, rely on a hand-designed action space, and meet problems due to the increasing size of the action space. In contrast, our method shows improved performance on typographic errors and also generalizes to issues with missing variable declaration errors by generating diverse fixes. 

In a recent work, Yasunaga and Liang \cite{yasunaga2020repair} proposed DrRepair to resolve the syntax error by introducing a program feedback graph. They connect the relevant symbols in the source code and the compile error messages and employ the graph neural network on top to model the process of the program repair. In this work, they rely on the compiler error messages which can be helpful, but it also limits the generality of the method. However, our proposed approach does not rely on additional information such as compiler error messages, and it resolves the errors by directly modeling the underlying distribution of the potential correct fixes.

\section{SampleFix: Generative Model for Diversified Code Fixes} 
Repairing the common program errors is a challenging task due to ambiguity in potential corrections and lack of representative data. Given a single erroneous program and a certain number of allowed changes, there are multiple ways to fix the program resulting in different styles and functionality. Without further information, the true, intended style and/or functionality remains unknown. In order to account for this inherent ambiguity, we propose a deep generative model to learn a distribution over potential fixes given the erroneous program -- in contrast to predicting a single fix. We frame this challenging learning problem as a conditional variational autoencoders (CVAE).
However, standard sampling procedures and limitations of datasets and their construction make learning and generation of diverse samples a challenge. We address this issue by a beam search decoding scheme in combination with a novel regularizer that encourages diversity of the samples in the embedding space of the CVAE.

\begin{figure*}[h] 
	\centering
	    \centering
		\includegraphics[width = 0.9\textwidth]{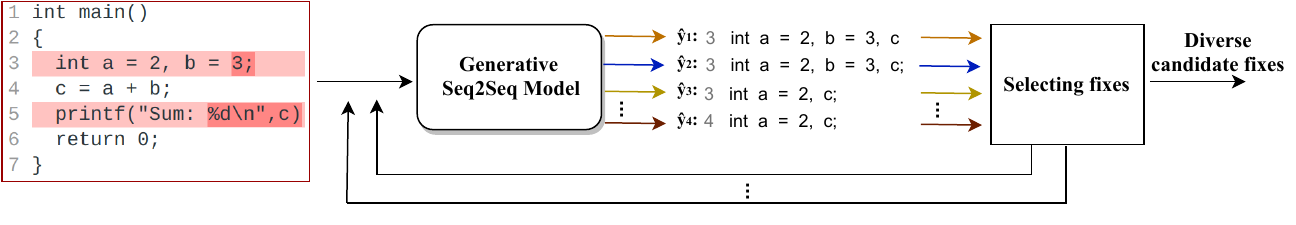}
	
	\caption{Overview of SampleFix at inference time, highlighting the generation of diverse fixes.}
	\label{fig:generative-model}
\end{figure*}
\figref \,\ref{fig:generative-model} provides an overview of our proposed approach at inference time. 
For a given erroneous program, the generative model draws $T$ intermediate, candidate fixes $\hat{y}$ from the learned conditional distribution. We use a compiler to select a subset of promising intermediate candidate fixes based on the number of remaining errors. This procedure is applied iteratively until arrive at a set of candidate fixes within the maximum number of prescribed changes. We then select a final set of candidate fixes that compile, have unique syntax according to our measure described below (Subsection ~\ref{subsec:selecting}).

In the following, we formulate our proposed generative model with the diversity regularizer and provide details of our training and inference process.  

\subsection{Conditional Variational Autoencoders for Generating Fixes}
Conditional Variational Autoencoders (CVAE) \cite{cvae15sohn}, model conditional distributions $p_{\theta}(\textbf{y} | \textbf{x})$ using latent variables  $\textbf{z}$. The conditioning introduced through $\textbf{z}$ enables the modelling of complex multi-modal distributions. As powerful transformations can be learned using neural networks, $\textbf{z}$ itself can have a simple distribution which allows for efficient sampling. This model allows for sampling from $p_{\theta}(\textbf{y} | \textbf{x})$ given an input sequence $\textbf{x}$, by first sampling latent variables $\hat{\textbf{z}}$ from the prior distribution $p({\textbf{z}})$. During training, amortized variational inference is used and the latent variables $\textbf{z}$ are learned using a recognition network $q_{\phi}(\textbf{z} | \textbf{x}, \textbf{y})$, parametrized by $\phi$.  In detail, the variational lower bound of the model (\equref \,\ref{eq:elbo}) is maximized,
\begin{equation}\label{eq:elbo}\tag{1}
\begin{aligned}
    \log(p(\textbf{y} | \textbf{x})) \geq \mathbb{E}_{q_{\phi}(\textbf{z} | \textbf{x}, \textbf{y})} &\log( p_{\theta}(\textbf{y} | \textbf{z}, \textbf{x}) ) \nonumber \\ &- D_\text{KL}({q_{\phi}(\textbf{z} | \textbf{x}, \textbf{y})},{p(\textbf{z} | \textbf{x})}).
  \end{aligned}
\end{equation}

Penalizing the divergence of $q_{\phi}(\textbf{z} | \textbf{x}, \textbf{y})$ to the prior in \equref \,\ref{eq:elbo} allows for sampling from the prior $p({\textbf{z}})$ during inference. In practice, the variational lower bound is estimated using Monte-Carlo integration,
\begin{equation}\label{eq:cvae}\tag{2}
\begin{aligned}
\hat{\mathcal{L}}_{\text{CVAE}} = \frac{1}{T}\sum\limits_{\text{i}=1}^{T} &\log( p_{\theta}(\textbf{y} | \hat{\textbf{z}}_{\text{i}}, \textbf{x}) ) 
 \\&- D_\text{KL}({q_{\phi}(\textbf{z} | \textbf{x}, \textbf{y})},{p(\textbf{z} | \textbf{x})})\enspace{.}
 \end{aligned}
\end{equation}

where, $\hat{\textbf{z}}_{\text{i}} \sim q_{\phi}(\textbf{z} | \textbf{x}, \textbf{y})$, and $T$ is the number of samples. We cast our model for resolving program errors in the Conditional Variational Autoencoder framework. Here, the input $\textbf{x}$ is the erroneous program and $\textbf{y}$ is the fix. 

However, the plain CVAE as described in \cite{cvae15sohn} suffers from diversity issues. Usually, the drawn samples do not reflect the true variance of the posterior $p(\textbf{y} | \textbf{x})$. This would amount to the correct fix potentially missing from our candidate fixes. To mitigate this problem, next we introduce an objective that aims to enhance the diversity of our candidate fixes. 

\subsection{Enabling Diverse Samples using a Best of Many Objective}
\label{subsec:samplefix}

Here, we introduce the diversity enhancing objective that we use. Casting our model in the Conditional Variational Autoencoder framework would enable us to sample a set of candidate fixes for a given erroneous program. However, the standard variational lower bound objective does not encourage diversity in the candidate fixes. This is because the average likelihood of the candidate fixes is considered. In detail, as the average likelihood is considered, all candidate fixes must explain the ``true'' fix in training set well. This discourages diversity and constrains the recognition network, which is already constrained to maintain a Gaussian latent variable distribution. In practice, the learned distribution fails to fully capture the variance of the true distribution. To encourage diversity, 
we employ "Many Samples" (MS) objective proposed by Bhattacharyya
et al. \cite{bhattacharyya2018accurate},
\begin{equation}\label{eq:ms}\tag{3}
    \begin{aligned}
    \hat{\mathcal{L}}_{\text{MS}} = \log \big(\frac{1}{T}\sum\limits_{\text{i}=1}^{T} & p_{\theta}(\textbf{y} | \hat{\textbf{z}}_{\text{i}}, \textbf{x})   \big) 
    \\&- D_\text{KL}({q_{\phi}(\textbf{z} | \textbf{x}, \textbf{y})},{p(\textbf{z} | \textbf{x})})\enspace{.}
    \end{aligned}
\end{equation}

In comparison to \equref \,\ref{eq:cvae}, this objective (\equref \,\ref{eq:ms}) encourages diversity in the model by allowing for multiple chances to draw highly likely candidate fixes. This enables the model to generate diverse candidate fixes, while maintaining high likelihood. In practice, due to numerical stability issues, 
we use "Best of Many Samples" (BMS) objective, which is an approximation of \ref{eq:ms}. This objective retains the diversity enhancing nature of \equref \,\ref{eq:ms} while being easy to train,

\begin{equation}\label{eq:bms}\tag{4}
    \begin{aligned}
    \hat{\mathcal{L}}_{\text{BMS}} = \max_{\text{i}} \big( &\log( p_{\theta}(\textbf{y} | \hat{\textbf{z}}_{\text{i}}, \textbf{x}) )  \big) 
    \\&- D_\text{KL}({q_{\phi}(\textbf{z} | \textbf{x}, \textbf{y})},{p(\textbf{z} | \textbf{x})})\enspace{.}
    \end{aligned}
\end{equation}

\subsection{\dssmaplefix: Encouraging Diversity with a Diversity-sensitive Regularizer}
\label{subsec:ds}
To increase the diversity using \equref \,\ref{eq:bms} we need to use a substantial number  of samples  during training. This is computationally prohibitive especially for large models, as memory requirements and computation time increases linearly in the number of such samples. On the other hand, for a small number of samples, the objective behaves similarly to the standard CVAE objective as the recognition network has fewer and fewer chances to draw highly likely samples/candidate fixes, thus limiting diversity. Therefore, in order to encourage the model to generate diverse fixes even with a limited number of samples, 
we propose a novel regularizer that aims to increase the distance between the two closest candidate fixes  (\equref \,\ref{eq:ds}). This
penalizes generating similar candidate fixes for a given erroneous program and thus encourages diversity in the set of candidate fixes. In comparison to \equref \,\ref{eq:bms}, we observe considerable gains even with the use of only $T=2$ candidate fixes. In detail, we maximize the following objective

\begin{equation}\label{eq:ds}\tag{5}
\begin{aligned}
    \hat{\mathcal{L}}_{\text{DS-BMS}} = \max_{\text{i}} \big( \log( p_{\theta}(\textbf{y} | \hat{\textbf{z}}_{\text{i}}, \textbf{x}) )  \big) 
    + \min_{\text{i},\text{j}} d(\hat{\textbf{y}}^{\text{i}},\hat{\textbf{y}}^{\text{j}}) \nonumber \\
    - D_\text{KL}({q_{\phi}(\textbf{z} | \textbf{x}, \textbf{y})},{p(\textbf{z} | \textbf{x})})\enspace{.}
\end{aligned}
\end{equation}

\myparagraph{Distance Metric}
\label{para:metric}
Here, we discuss the distance metric $d$ in \equref \,\ref{eq:ds}. Note, that the samples $\left\{ \hat{\textbf{y}}^{\text{i}},\hat{\textbf{y}}^{\text{j}} \right\}$ can be of different lengths. Therefore, we first pad the shorter sample to equalize lengths. Empirically, we find that the Euclidean distance performs best. This is mainly because, in practice, Euclidean distance is easier to optimize.

\subsection{Beam Search Decoding for Generating Fixes}
Beam search decoding is a classical model to generate multiple outputs from a sequence-to-sequence model \cite{diverse_accurate,deshpande2019fast}. Given the distributions $p_{\theta}(\textbf{y} | \textbf{x})$ of a sequence-to-sequence model we can generate multiple outputs by unrolling the model in time and keeping the top-K tokens at each time step, where $K$ is the beam width. In our generative model, we employ beam search algorithm to sample multiple fixes. In detail,  we decode with beam width of size $K$ for each sample $\textbf{z}$ and in total for $T$ samples from $p(\textbf{z})$. We set $T = 100$ during inference.

\subsection{Selecting Diverse Candidate Fixes}
\label{subsec:selecting}

We extend the iterative repair procedure introduced by Gupta et al. \cite{Gupta2017DeepFixFC} in the context of our proposed generative model, where the iterative procedure now leverages multiple candidate fixes.
Given an erroneous program, the generative model outputs $T$ candidate fixes. Each fix contains a potential erroneous line with the corresponding fix. So in each iteration we only edit one line of the given program. To select the best fixes, we take the candidate fixes and the input erroneous program, reconcile them to create $T$ updated programs. We evaluate these fixes using a compiler, and select up to the best $N$ fixes, where $N \leq T$. We only select the unique fixes which do not introduce any additional error messages. In the next iterations, we feed up to $N$ programs back to the model. These programs are updated based on the selected fixes of the previous iteration. We keep up to $N$ programs with the lower number of error messages over the iterations. At the end of the repairing procedure, we obtain multiple potential candidate fixes. In the experiments where we are interested in a single repaired program, we pick the best fix with the highest probability score according to our deep generative model.

\subsection{Model Architecture and Implementation Details}
	
\begin{wrapfigure}[11]{R}{0.45\linewidth}
\vspace{-.8cm}
\centering
    \includegraphics[width=\linewidth]{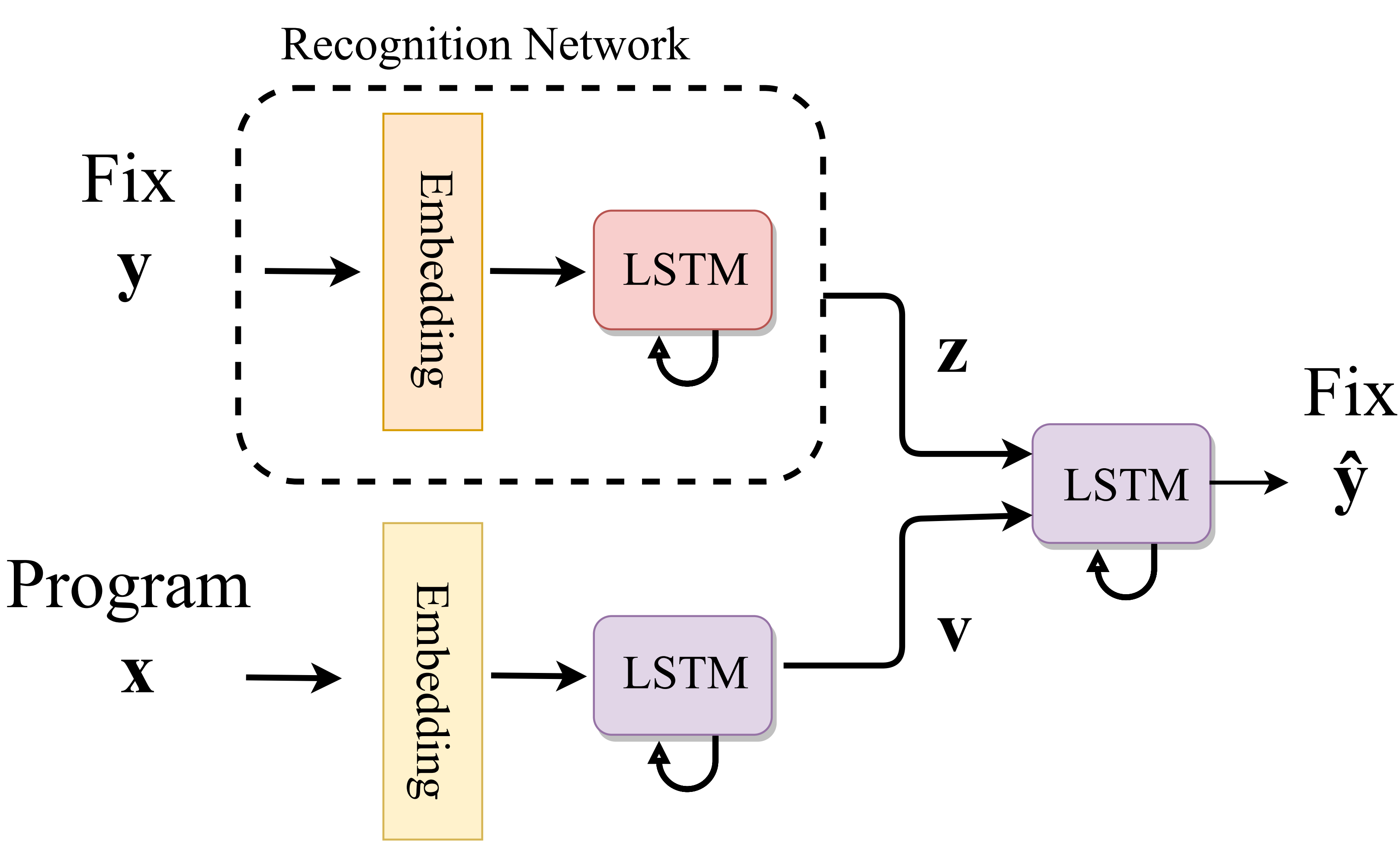}
    \caption{Overview of network architecture.}
\label{fig:net}
\end{wrapfigure}

To ensure a fair comparison, our generative model is based on the sequence-to-sequence architecture, similar to Gupta et al. \cite{Gupta2017DeepFixFC}. \figref \,\ref{fig:net} shows the architecture of our approach in detail. Note that the recognition network is available to encode the fixes to latent variables $\textbf{z}$ only during training.
All of the networks in our framework consists of 4-layers of LSTM cells with 300 units. The network is optimized using Adam optimizer \cite{kingma15adam} with the default setting.
We use $T = 2$ samples to train our models, and $T = 100$ samples during inference. To process the program through the networks, we tokenize the programs similar to the setting used by Gupta et al. \cite{Gupta2017DeepFixFC}.

During inference, the conditioning erroneous program $\textbf{x}$ is input to the encoder, which encodes the program to the vector $\textbf{v}$. To generate multiple fixes using our decoder, the code vector $\textbf{v}$ along with a sample of $\textbf{z}$ from the prior $p(\textbf{z})$ is input to the decoder. For simplicity, we use a standard Gaussian $\mathcal{N}(0,\textbf{I})$  prior, although more complex priors can be easily leveraged. The decoder is unrolled in time and output logits ($p_{\theta}(\textbf{y} | \hat{\textbf{z}}_{\text{i}}, \textbf{x})$).

\section{Experiments}
\label{experiments}
We evaluate our approach on the task of repairing common programming errors. %
We %
     evaluate the diversity and accuracy of our sampled error corrections as well as 
     compare our proposed method with the state of the art.
\begin{wraptable}[14]{R}{0.46\linewidth}
\vspace{-.5cm}
\fontsize{8.5}{10.5}\selectfont
\begin{center}
\caption{Results of performance comparison of DeepFix, Beam search (BS), \samplefix \,,and \dssmaplefix\, on synthetic data. Typo, Miss Dec, and All refer to typographic, missing variable declarations, and all of the errors respectively.}
\label{table:seeded}
\end{center}

\begin{center}
\begin{tabular}{lccc}
\toprule
Models & Typo & Miss Dec & All\\
\cmidrule(lr){1-1} \cmidrule(lr){2-4} 
DeepFix &   84.7\% &    78.8\% & 82.0\%\\
Beam search (BS) &    91.8\% &   89.5\% & 90.7\%\\
\samplefix &    86.8\% &   86.5\% & 86.6\%\\
\dssmaplefix &  95.6\% &    88.1\% & 92.2\% \\
\bottomrule
\end{tabular}
\end{center}

\end{wraptable}

\begin{savenotes}
\begin{table*}[h]

 \fontsize{8.5}{10.5}\selectfont
 
\begin{center}
\caption{Results of performance comparison of DeepFix, RLAssist, DrRepair, Beam search (BS), \samplefix \,, \dssmaplefix, and \dssmaplefix\,+ BS. Typo, Miss Dec, and All refer to typographic, missing variable declarations, and all of the error messages respectively. Speed denotes computational time for sampling 100 fixes. \faCheck denotes successfully compiled programs, while \faBug \, refers to resolved error messages.}
\label{table:acc}
\end{center}
\begin{center}
\begin{tabular}{lccccccc}
\toprule
Models   &  \multicolumn{2}{c}{Typo }& \multicolumn{2}{c}{Miss Dec} & \multicolumn{2}{c}{All} & Speed (s)\\ \cmidrule(lr){1-1}\cmidrule(lr){2-3}\cmidrule(lr){4-5}\cmidrule(lr){6-7}\cmidrule(lr){8-8}
             & \hspace{5pt}\faCheck    & \hspace{5pt}\faBug & \hspace{5pt}\faCheck  & \hspace{5pt}\faBug & \hspace{5pt}\faCheck   & \hspace{5pt}\faBug \\
DeepFix \cite{Gupta2017DeepFixFC} &  23.3\% &   30.8\%       & 10.1\%         &     12.9\%     &      33.4\%  &       40.8\%    & -\\ 
\textit{RLAssist} \cite{gupta2019RLAssist} &  \textit{26.6}\%       &      \textit{39.7}\%     &    -      &        -     &   -     &         -    & -\\ 
DrRepair \cite{yasunaga2020repair} &  -     &      -     &    -      &        -     &   34.0\%     &      -    & -\\ 
Beam search (BS) &  25.9\%        &     42.2\%     &   \textbf{20.3\%}        &    47.0\%     &   44.7\%     &  63.9\% & 4.82\\

\samplefix  &     24.8\%     &     38.8\%     &  16.1\%        &      22.8\%     &     40.9\%   &   56.3\%  & 0.88 \\ 
\dssmaplefix &  27.7\%        &      40.9\%     &   16.7\%       &     24.7\%     &   44.4\%     &   61.0\%     & 0.88 \\

\dssmaplefix \,+ BS & \textbf{27.8\%}        &       \textbf{45.6\%}     &   19.2\%       &    \textbf{47.9\%}     &   \textbf{45.2\%}     &    \textbf{65.2\%}  & 1.17
\\\bottomrule
\end{tabular}

\end{center}
\end{table*}
\end{savenotes}

\subsection{Dataset} 
We use the dataset published by Gupta et al. \cite{Gupta2017DeepFixFC} as it's sizable and includes real-world data. It contains C programs written by students in an introductory programming course. The dataset consists of 93 different tasks that were written by students in an introductory programming course. 
The programs were collected using a web-based system \cite{das16tutorin}. These programs have token lengths in the range $\left[75,450\right]$, and 
contain typographic and missing variable declaration errors. To tokenize the programs and generate training and test data  different type of tokens, such as types, keywords, special characters, functions, literals and variables are used. The dataset contains two sets of data which are called synthetic and real-world data. The synthetic data contains the erroneous programs which are synthesized by mutating correct programs written by students. The real-world data contains 6975 erroneous programs with 16766 error messages.
\subsection{Evaluation}
\label{subsec:eval}

We evaluate our approach on synthetic and real-world data. To evaluate our approach on the synthetic test set we randomly select 20k pairs. This data contains pairs of erroneous programs with the intended fixes. To evaluate our approach on real-world data we use a real-world set of erroneous programs. Unlike synthetic test set, we don't have access to the intended fix(es) in the real-world data. However, we can check the correctness of the program using the evaluator (compiler). Following the prior work, we train two networks, one for typographic errors and another to fix missing variables declaration errors. Note that there might be an overlap between the error resolved by the network for typographic errors and the network for missing variables declaration errors, so we also provide the overall results of the resolved error messages. 

\subsubsection{Synthetic Data.} 
\tableref \,\ref{table:seeded} shows the comparison of our proposed approaches, Beam search (BS), \,\samplefix \, and \dssmaplefix, with DeepFix \cite{Gupta2017DeepFixFC} on the synthetic data in the first iteration. In this table (\tableref \,\ref{table:seeded}), we can see that our approaches outperform DeepFix in generating intended fixes for the typographic and missing variable declaration errors. Beam search (BS), \,\samplefix \, and \dssmaplefix \,generate 90.7\%, 86.6\%, and 92.2\% of the intended fixes 
respectively. 
\subsubsection{Real-World Data.}  In \tableref \,\ref{table:acc} we compare our approaches, with state-of-the-art approaches (DeepFix \cite{Gupta2017DeepFixFC}, RLAssist \cite{gupta2019RLAssist}, and DrRepair \cite{yasunaga2020repair}) on the real-world data. In our experiments (\tableref \,\ref{table:acc}) we show the performance of beam search decoding, CVAEs (\samplefix), and our proposed diversity-sensitive regularizer (\dssmaplefix). Furthermore, we show that \dssmaplefix\, can still take advantage of beam search algorithm (\dssmaplefix\,+~BS). To do that, for each sample $z$ we decode with beam width of size 5, and to sample 100 fixes we draw 20 samples from $p(z)$. We also provide the sampling speed in terms of sampling 100 fixes for a given program using an average over 100 runs. The running time results show that CVAE-based models are at least 4x faster than beam search in sampling the fixes. In this experiment, we feed the programs up to 5 iterations.

\tableref \,\ref{table:acc} shows that our approaches outperform DeepFix \cite{Gupta2017DeepFixFC}, RLAssist \cite{gupta2019RLAssist}, and DrRepair \cite{yasunaga2020repair} in resolving the error messages. This shows that generating multiple diverse fixes can lead to substantial improvement in performance. Beam search, \samplefix, \dssmaplefix, and \dssmaplefix \,+ BS resolve 63.9\%, 56.3\%, 61.0\%, and 65.2\% of the error messages respectively.
Overall, our \dssmaplefix \,+ BS is able to resolve all compile-time errors of the $45.2\%$ of the programs - around $12\%$ points improvement over DeepFix and $11\%$ points improvement over DrRepair. Furthermore, the performance advantage of \dssmaplefix \, over \samplefix \, shows the effectiveness of our novel regularizer.

Note that DrRepair \cite{yasunaga2020repair} has achieved further improvements by relying on the compiler. While utilizing the compiler output seems to be beneficial, it also limits the generality of the approach.
For a fair comparison, we report the performance of DrRepair without the compiler output, but consider informing our model by the compiler output an interesting avenue of future work.

\begin{figure*}[h] 
	\centering
	    \centering
		\includegraphics[width = 0.8\textwidth]{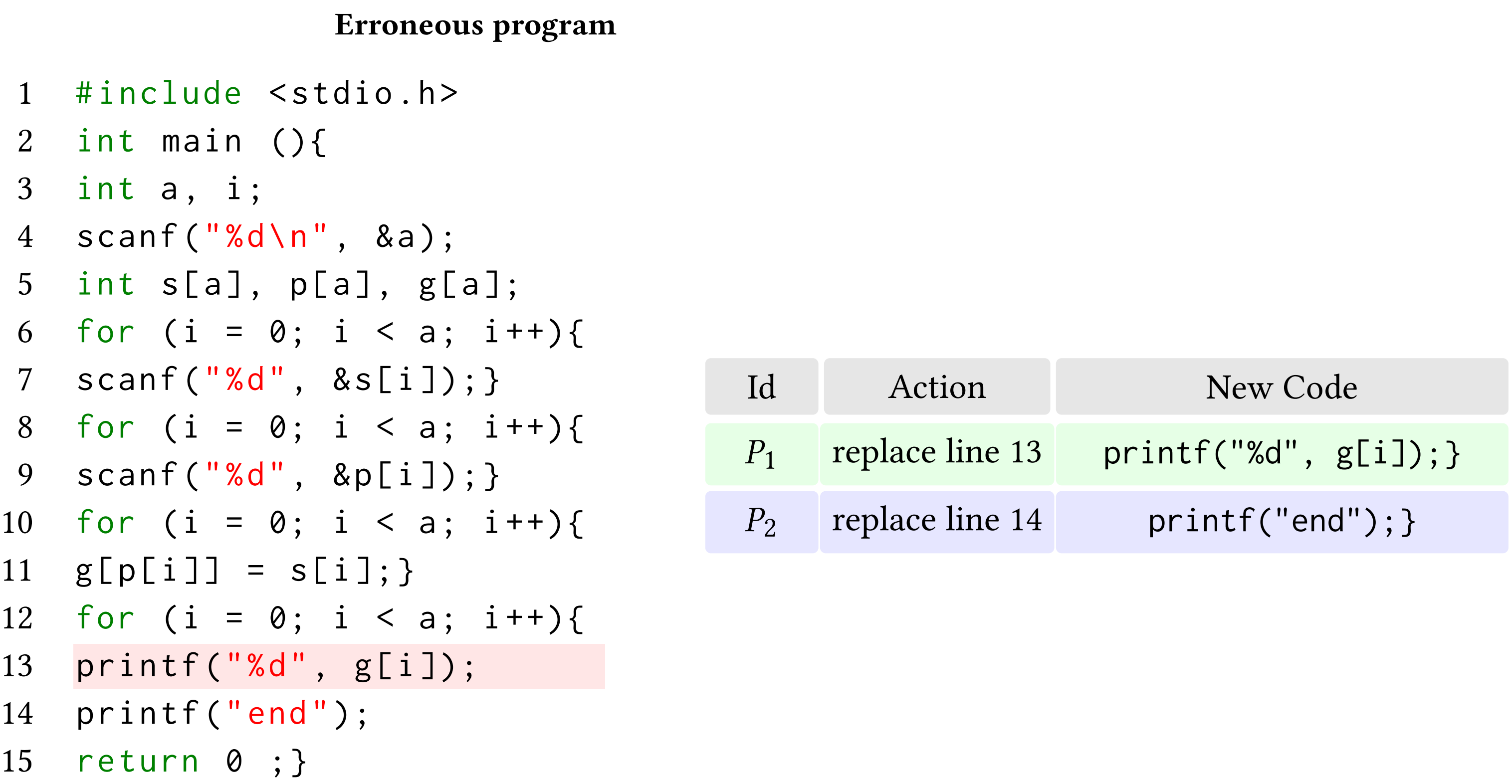}
	
	\caption{An example illustrating that our DS-SampleFix can generate diverse fixes. Left: Example of a program with a typographic error. The error, i.e., missing bracket, is highlighted at line 13. Right: Our DS-SampleFix proposes multiple fixes for the given error (line number with the corresponding fix), highlighting the ability of DS-SampleFix to generate diverse and accurate fixes.}
	\label{fig:example1}
\end{figure*}

\subsubsection{Qualitative Example.} 
We illustrate diverse fixes generated by our DS-SampleFix in \figref \,\ref{fig:example1} using a code example with typographic errors, with the corresponding two output samples of \dssmaplefix. In the examples given in \figref \,\ref{fig:example1}, there is a missing closing curly bracket after line 13. We can see that DS-SampleFix generates multiple correct fixes to resolve the error in the given program.
This indicates that our approach is capable of handling the inherent ambiguity and uncertainty in predicting fixes for the erroneous programs. The two fixes in \figref \,\ref{fig:example1} are unique and compileable fixes that implement different functionalities for the given erroneous program. Note that generating multiple diverse fixes gives the programmers the opportunity of choosing the desired fix(es) among the compileable ones, based on their intention.

\subsubsection{Generating Functionally Diverse Programs.}
Given an erroneous program, our approach can generate multiple potential fixes that result in a successful compilation. Since we do not have access to the user's intention, it is desirable to suggest multiple potential fixes with diverse functionalities. Here, we evaluate our approach in generating multiple programs with different functionalities. 

In order to assess different functionalities, we use the following approach based on tests.
The dataset of Gupta et al. \cite{Gupta2017DeepFixFC} consists of 93 different tasks. The description of each task, including the input/output format, is provided in the dataset. Based on the input/output format, we can provide input examples for each task. To measure the diversity in functionality of the programs in each task, we generate 10 input examples. For instance, given a group of programs for a specific task, we can run each program using the input examples and get the outputs. We consider two programs to have different functionalities if they return different outputs given the same input example(s). 

In order to generate multiple programs we use our iterative selecting strategy (Subsection ~\ref{subsec:selecting}).
In each iteration, we keep up to $N$ programs with the less number of error messages over the iterations. At the end of the repairing procedure, we obtain multiple repaired programs. As discussed (\figref \,\ref{fig:bubbles}), a subset of these programs will successfully compile. In this experiment, we use the real-world test set, and we set $N = 50$ as this number is large enough to allow us to study the diversity of the fixes, without incurring an unnecessarily large load on our infrastructure. Our goals in the remaining of this section are: \begin{enumerate*}
    \item For each erroneous program, to measure the number of  generated unique fixes that successfully compile.
    \item For each erroneous program, to measure the number of generated programs with different functionalities.
\end{enumerate*}

\begin{figure*}
	\centering
	    \centering
	    
	    \begin{subfigure}[b]{0.49\textwidth}
	    \centering
		\includegraphics[width=\linewidth]{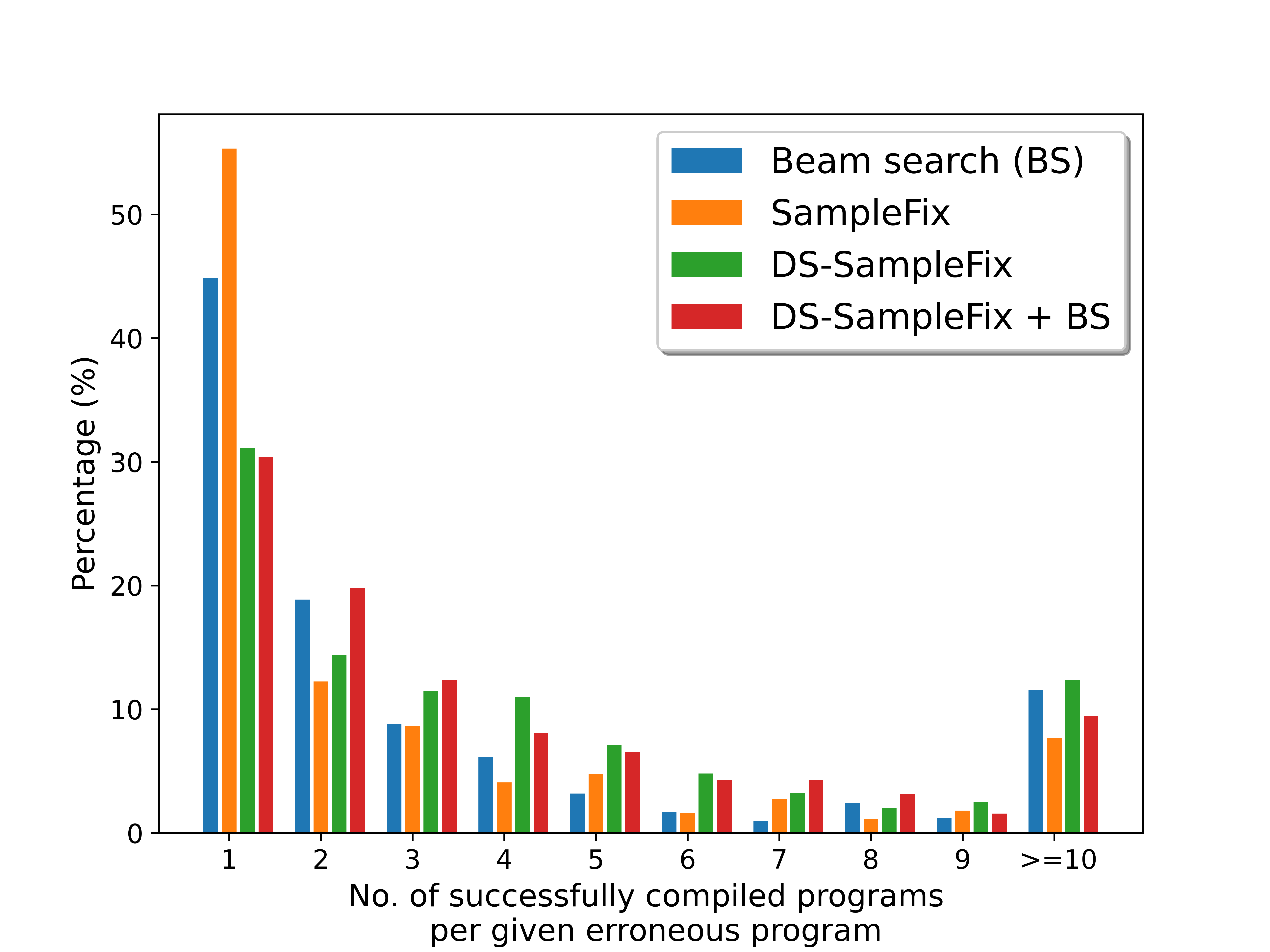}
		\caption{Diversity of the generated programs.}
		\label{fig:diff-prog}
		\vspace{.4cm}
    	\end{subfigure} %
    	\hfill
    	\begin{subfigure}[b]{0.49\textwidth}
    	    \centering
    		\includegraphics[width=\linewidth, ]{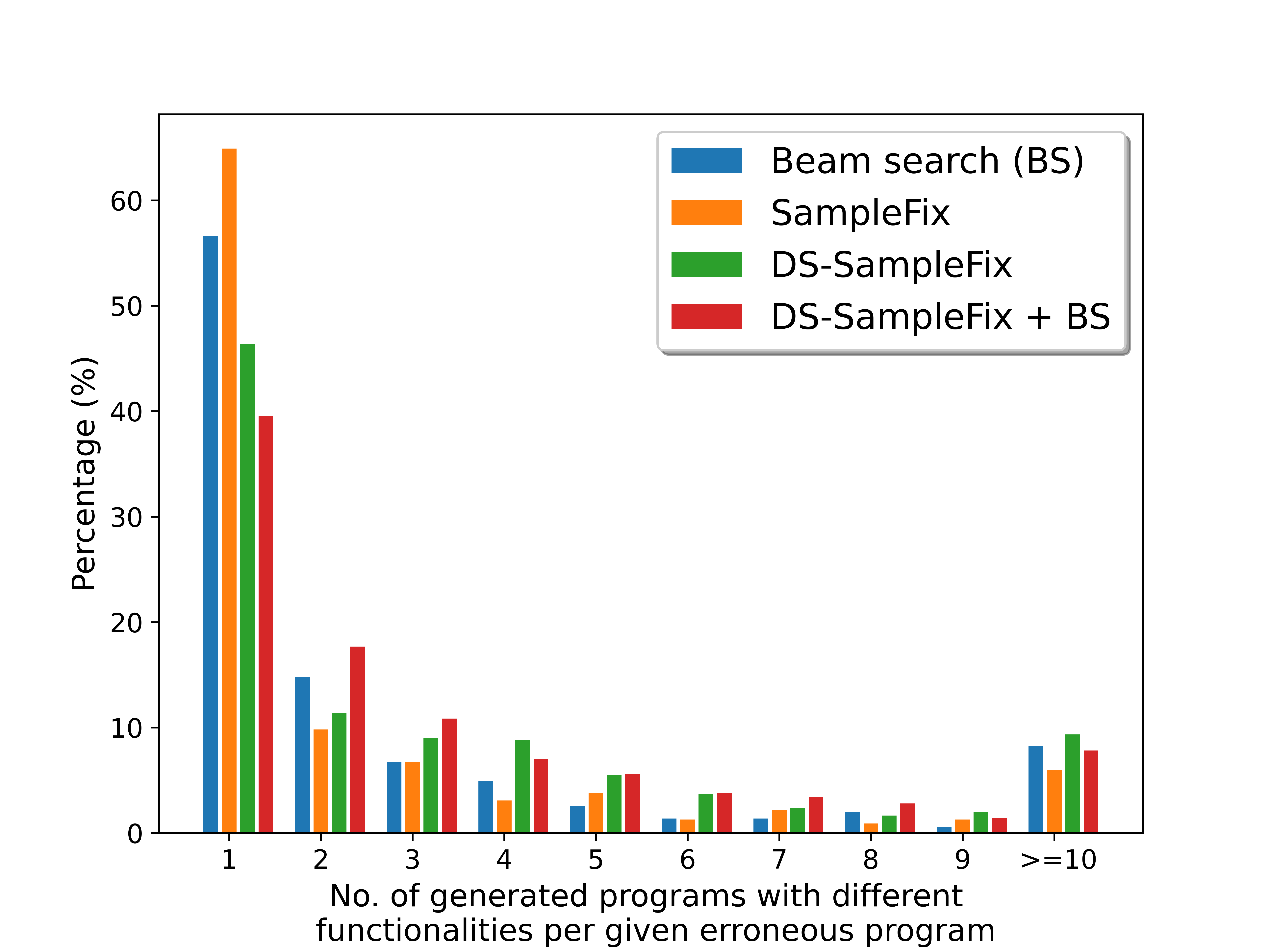} 
    		\caption{Diversity of the functionality of the generated programs.}
    		\label{fig:diff-func}
    	\end{subfigure} 
	
	\caption{The results show the performance of Beam search (BS), \samplefix\,, \dssmaplefix\,, and \dssmaplefix\,+ BS. (a) Percentage of the number of the generated successfully compiled, unique programs for the given erroneous programs. (b) Percentage of the successfully compiled programs with different functionalities for the given erroneous programs. }
	\label{fig:func}
	\vspace{-.5cm}
\end{figure*}

\begin{wraptable}[17]{R}{0.55\linewidth}
\vspace{-1.25cm}
\fontsize{8.5}{10.5}\selectfont
\begin{center}
\caption{Results of performance comparison of Beam Search (BS), \samplefix \,, \dssmaplefix \,, and \dssmaplefix\,+BS on generating diverse programs.  Diverse Prog refers to the percentage of cases where the models generate at least two or more successfully compiled unique programs. Diverse Func denotes the percentage of cases where the models generate at least two or more programs with different functionalities.}
\label{table:func}
\end{center}

\begin{center}
\begin{tabular}{lcc}
\toprule
Models & Diverse Prog & Diverse Func\\
\cmidrule(lr){1-1} \cmidrule(lr){2-3} 
Beam search &    55.6\% &   45.1\%\\
\samplefix &    44.6\% &   34.9\%\\
\dssmaplefix &  68.8\% &    53.4\%\\
\dssmaplefix\,+ BS &   \textbf{69.5}\% &    \textbf{60.4}\%\\

\bottomrule
\end{tabular}
\end{center}

\end{wraptable}

\figref \,\ref{fig:diff-prog} and \figref \,\ref{fig:diff-func} show the syntactic diversity of the generated programs, and the diversity in functionality of these programs, respectively. In \figref \,\ref{fig:diff-prog}
we show the percentage of the successfully compiled programs with unique fixes for a given erroneous program. The x-axis refers to the number of generated and successfully compiled unique programs, and y-axis to the percentage of repaired programs for which these many unique fixes were generated. For example, for almost 20\% of the repaired programs, \dssmaplefix\,+~BS generates two unique fixes. Overall, we can see that \dssmaplefix\, and \dssmaplefix\,~+~BS generate more diverse programs in comparison to the other approaches. 

\figref \,\ref{fig:diff-func} shows the percentage of the successfully compiled programs with different functionalities, for a given erroneous program. Here, the x-axis refers to the number of the generated functionally different programs, and y-axis refers to the percentage of erroneous programs with at least one fix, for which we could generate that many diverse fixes. One can observe that in many cases, e.g., up to 60\% of the times for \samplefix, the methods generate programs corresponding to a single functionality. However, in many other cases they generate functionally diverse fixes. For example, in almost 10\% of the cases, \dssmaplefix\, generate 10 or more fixes with different functionalities.
In \figref \,\ref{fig:diff-func} we can see that all of the approaches have higher percentage for generating program with the same functionality in comparison to the results in \figref \,\ref{fig:diff-prog}. This indicates that for some of the given erroneous programs, we generate multiple unique programs with approximately the same functionality. These results show that \dssmaplefix\, and \dssmaplefix\,+ BS  generate programs with more diverse functionalities in comparison to the other approaches.

In \tableref \,\ref{table:func} we compare the performance of our approaches in generating diverse programs and functionalities. We provide results for all of our four approaches, i.e., Beam search (BS), \samplefix\,, \dssmaplefix\,, and \dssmaplefix\,+ BS. We consider that an approach can generate diverse programs if it can produce two or more successfully compiled, unique programs for a given erroneous program. Similarly, we say that the approach produces functionally diverse programs if it can generate two or more programs with observable differences in functionality for a given erroneous program. Here we consider the percentage out of the total number of erroneous programs for which the model generates at least one successfully compiled program. The results of this table show that our \dssmaplefix\,+ BS approach generates programs with more diverse functionalities in comparison to the other approaches.

\section{Conclusion}
We propose a novel approach to correct common programming errors. We recognize and model the inherent ambiguity and uncertainty when predicting multiple fixes. In contrast to previous approaches, our approach is able to learn the distribution over candidate fixes rather than the most likely fix. 
We achieve increased diversity of the sampled fixes by a novel diversity-sensitive regularizer. We show that our approach is capable of generating multiple diverse fixes with different functionalities. Furthermore, our evaluations on synthetic and real-world data show improvements over state-of-the-art methods.

\bibliographystyle{splncs04}
\bibliography{samplefix}
\end{document}